# Vesta and Ceres: crossing the history of the Solar System

Coradini A. [1,*], Turrini D. [1], Federico C. [2], Magni G. [3]

[1] *Institute for Physics of Interplanetary Space, INAF, Rome, Italy*
[2] *Department of Earth Sciences, University of Perugia, Perugia, Italy*
[3] *Institute for Space Astrophysics and Cosmic Physics, INAF, Rome, Italy*
*Corresponding author: angioletta.coradini@ifsi-roma.inaf.it

**Abstract**

*The evolution of the Solar System can be schematically divided into three different phases: the Solar Nebula, the Primordial Solar System and the Modern Solar System. These three periods were characterized by very different conditions, both from the point of view of the physical conditions and from that of the processes there were acting through them. Across the Solar Nebula phase, planetesimals and planetary embryos were forming and differentiating due to the decay of short-lived radionuclides. At the same time, giant planets formed their cores and accreted the nebular gas to reach their present masses. After the gas dispersal, the Primordial Solar System began its evolution. In the inner Solar System, planetary embryos formed the terrestrial planets and, in combination with the gravitational perturbations of the giant planets, depleted the residual population of planetesimals. In the outer Solar System, giant planets underwent a violent, chaotic phase of orbital rearrangement which caused the Late Heavy Bombardment. Then the rapid and fierce evolution of the young Solar System left place to the more regular secular evolution of the Modern Solar System. Vesta, through its connection with HED meteorites, and plausibly Ceres too were between the first bodies to form in the history of the Solar System. Here we discuss the timescale of their formation and evolution and how they would have been affected by their passage through the different phases of the history of the Solar System, in order to draw a reference framework to interpret the data that Dawn mission will supply on them.*

## 1. Introduction

Vesta and Ceres, the two targets of Dawn mission, are among the largest members of the population of bodies which compose the Main Asteroid Belt, a relic of the protoplanetary disk that orbited the young Sun while the Solar System was shaping itself to its present form. From the point of view of their composition and internal structure, the rocky Vesta and ice-rich Ceres are extremely different. While distinct, however, Vesta and Ceres have two features in common. First, both asteroids formed at the beginning of the life of the Solar System. Second, theoretical models suggest that both bodies underwent a significant thermal and collisional evolution since their formation. The former feature implies that both bodies could be used to probe the more ancient and less understood epochs of the history of the Solar System. The latter, however, has as a consequence that most primordial features were removed across the subsequent 4.5 Ga of evolution.

In order to interpret correctly the data that will be obtained by the Dawn mission and to shed new light on the formation and evolution of the Solar System, we need to possess a reference scheme in which the data can be included. This chapter aims to describe such reference scheme for what it concerns the roles of Vesta and Ceres in understanding the formation and evolution of the Solar System. A complementary approach is supplied in the chapter by O'Brien & Sykes (2011), where Vesta and Ceres are discussed in the context of the evolution of the Main Asteroid Belt.

As we anticipated, it is likely that both Vesta and Ceres underwent to an extensive thermal evolution. At present, observational evidences and constrains about thermal evolution are available only for Vesta through its connection to HED (Howardites, Eucrites and Diogenites) meteorites, as we will discuss later. The same is not true for Ceres, for which we presently lack this kind of connection. The dwarf planet, however, can still play an important role to constrain the evolution of the early Solar System and the efficiencies of both the planetary accretion and depletion processes, as will be discussed in Sect. 2.1 and 2.2.

Can we find a record of the processes of planetary formation and differentiation on Vesta? We will try to evaluate here if we can answer to this question now, and how our knowledge will be improved after the Dawn mission. At present, our understanding of Vesta is based on a very strong assumption that is not yet completely proved, i.e. that the HED meteorites originated on Vesta. Differentiated meteorites provide a record of asteroidal melting, and age constraints from short-lived $^{182}$Hf-$^{182}$W, $^{60}$Fe-$^{60}$Ni, $^{53}$Mn-$^{53}$Cr, and $^{26}$Al-$^{26}$Mg isotopic systems suggest that planetesimals differentiation occurred within 10 Ma of the lifetime of the Solar System (see e.g. [Shukolyukov & Lugmair 1993](#); [Lugmair & Shukolyukov 1998](#); [Srinivasan et al. 1999](#); [Yin et al. 2002](#); [Kleine et al. 2002](#); [Quitte & Birck 2004](#); [Kleine et al. 2005](#); [Baker et al. 2005](#); [Bizzarro et al. 2005](#); [Markowsky et al. 2006](#); [Yang et al. 2007](#) and [Sect. 2.1.2](#) for further discussion). The energy sources that were responsible for the differentiation of Vesta were most likely the short-lived radioactive elements. This assumption is broadly consistent with the relatively young mineral ages obtained for differentiated meteorites by the differentiation due to short-lived radionuclides as $^{26}$Al and $^{60}$Fe and it implies that the differentiation of Vesta took place due to the decay of the $^{26}$Al nuclide, the principal heat source postulated to have induced planetesimal melting in the young Solar System ([Urey 1955](#)). However, as we will describe later, there are several discussions related to the original amount of short lived radioactive elements as well as to the time in which they were incorporated in the growing Vesta. The indetermination is as large as a few million of years, based on different dating methods and sampled pristine material ([Amelin et al. 2010](#)). Moreover, we have to take into account of the delay time between the generation and possible differentiation of original planetesimals and the formation of Vesta. Such considerations suggest that Vesta didn't undergo to an extensive surface melting, even if it was characterized by more than one period of large scale differentiation. Given this scenario, and the fact that the hypothetical Vestian material is very old, it is reasonable to search on the surface of Vesta records of different bombardment episodes that have characterized the primordial phases of the evolution of the Solar System.

## 2. Vesta, Ceres and the history of the Solar System

As we anticipated, the investigation of Vesta and Ceres will not simply allow us to gather information on the physics of the accretion and differentiation processes that took place in the Main Asteroid Belt but will also allow us to glimpse into the very early history of the Solar System. Schematically, the history of the Solar System can be viewed as composed by three different phases, each characterized by different physical processes and different durations: Solar Nebula, Primordial Solar System and Modern Solar System. In the following sections, we will describe in detail each of these three phases and the implications of the observations of Dawn mission for their comprehension.

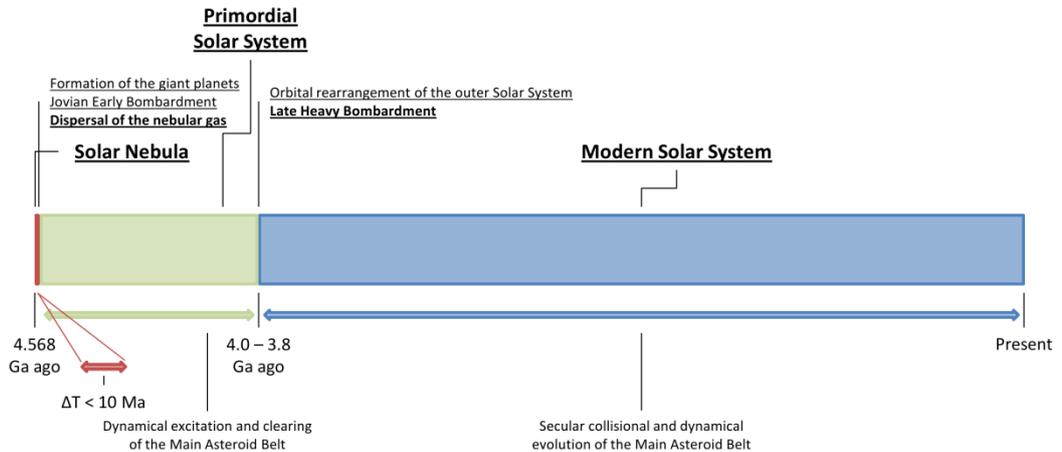

Figure 1: timeline of the evolution of the Solar System from the condensation of the first solids (Ca-Al-rich inclusions, CAIs in the following) to present. The three phases of the evolution of the Solar System (Solar Nebula, Primordial Solar System and Modern Solar System) are reported in different colors (respectively red, green and blue). The main events marking the transitions between different phases (i.e. the dispersal of the nebular gas and the orbital rearrangement of the outer Solar System) are marked in bold fonts.

## *2.1 Solar Nebula*

The first phase in the lifetime of the Solar System is that of the Solar Nebula (see Figure 1): during this phase, the Solar System is constituted by a circumsolar disk of gas and dust particles where planetesimals and planetary embryos are forming (see e.g. De Pater & Lissauer 2001; Bertotti, Farinella & Vokrouhlický 2003). The beginning of the Solar Nebula phase, which sets the age of the Solar System, is conventionally assumed to coincide with the condensation of the first solids in the circumsolar disk (ibid). The end of the Solar Nebula phase is instead marked by the dispersal of the nebular gas due to the Sun entering the T Tauri phase (ibid). At the end of the Solar Nebula phase, the Solar System is composed by a protoplanetary disk populated by planetesimals, planetary embryos and the four giant planets. The giant planets should have formed in the Solar Nebula since the nebular gas represents the source material for both the massive gaseous envelopes of Jupiter and Saturn and the limited ones of Uranus and Neptune (ibid). The formation of Jupiter, particularly across the final phase of gas accretion, is possibly the main event of this stage in the evolution of the Solar System. According to Lissauer et al. (2009), the gas-accretion time of Jupiter varies between several $10^4$ to a few $10^5$ years. Coradini, Magni & Turrini (2010) measured the gas-accretion time-scales for Jupiter and Saturn, which vary between a few $10^3$ to about $10^5$ years depending on the physical parameters of the Solar Nebula. The agreement between these values indicates that the time over which Jupiter accreted its gaseous envelope (i.e. about 1 Ma) is of the same order of magnitude of the accretion and differentiation timescales of the primordial planetesimals (see Scott 2006, 2007). Across the accretion of its gaseous envelope, moreover, theoretical models indicate that Jupiter should have migrated due to disk-planet interaction on a timescale of $10^5$ years (see e.g. Papaloizou et al. 2007 and references therein). The rapid changes in the gravitational potential across the circumsolar disk due to Jupiter's accretion and migration should have strongly affected the mass and velocity distribution of all the bodies present at that time in the Solar Nebula (see e.g. Coradini, Magni & Turrini 2010; Turrini, Magni & Coradini 2011).

Quantitative information on the age and the duration of the Solar Nebula phase is supplied respectively by meteoritic studies and astronomical observations. The radiometric ages of chondrites, achondrites and differentiated meteorites indicates that the first solids to form were the Ca-Al-rich inclusions (Amelin et al. 2002), CAIs in the following, whose condensation time dates back to 4567.2 Ma ago (ibid) or, as recent results suggest, even 4568.2 Ma ago (Bouvier & Wadhwa 2010). The duration of the Solar Nebula phase is indirectly constrained through the astronomical observations of circumstellar disks, which indicate that their average lifetime is about 3 Ma, with the range of observed values spanning between 1 - 10 Ma (Haisch, Lada & Lada 2001; Jayawardhana et al. 2006, Meyer 2008). Disk accretion ceases or dips below measurable levels by 10 Ma in the vast majority of low-mass stars (Jayawardhana et al. 2006). While a few disks can continue to accrete for up to ~10 Ma, disks accreting beyond this timescale are rather rare (ibid).

## 2.1.1 Planetesimal accretion and differentiation

Across this ΔT < 10 Ma timespan, as we previously mentioned, planetary accretion was acting in the Solar Nebula to form the planetesimals, the planetary embryos and the giant planets that will later populate the Primordial Solar System (Safronov 1969). However, it is difficult to gather information on such a remote time since most features of this ancient period were cancelled by the subsequent 4.5 Ga of dynamical and collisional evolution. Presently, the radiometric studies of meteoritic materials represent our best probe into the processes that were acting at the time of the Solar Nebula. According to radiometric ages chondrules, once thought to represent the oldest material that solidified in the Solar Nebula, formed about 1 - 3 Ma later than CAIs (Amelin et al. 2002, Connelly et al. 2008) while differentiated bodies generally appeared in the next few million years after the formation of chondrules (see Scott 2007 and references therein). Meteoritic evidences, however, suggest that in some cases differentiation of planetesimals took place extremely early in the history of the Solar System, i.e. about 1-2 Ma after the formation of CAIs (Baker et al. 2005; Bizzarro et al. 2005). Such primordial differentiation was due to the presence of short-lived radionuclides, mainly $^{26}$Al and $^{60}$Fe (see e.g. Bizzarro et al. 2005) in bodies larger than 20 - 30 km in radius (Scott 2007). In particular, the results by Yang, Goldstein & Scott (2007) obtained in studying the iron meteorites from the *IV A* group suggest that these meteorites originated from a parent body that was about 300 km in diameter and lacked an insulating mantle. These authors explained such anomalous composition of the parent body through the removal of the silicate-rich mantle from a differentiated parent body whose original size was about $10^3$ km in diameter (ibid).

All these results collectively imply that the processes of planetary accretion and differentiation started at the very beginning of the history of the Solar System and that a first generation of hundreds-of-km-wide bodies formed and differentiated in the first few Ma, contemporarily to the formation of chondrules (see Scott 2007 for a more detailed discussion). Indirect confirmations to the early formation of the planetesimals come from the study of debris disks around nearby stars. Dust forms naturally as a by-product of planetary formation (Kenyon & Bromley 2002), when massive (about 1000 km in diameter or greater) objects form and cause a collisional cascade in smaller objects that grinds them to dust (Kenyon & Bromley 2002, 2008, 2010). Models of debris disks applied to the observations of circumstellar dust disks, like e.g. the ones orbiting β Pictoris (Hahn 2011) or HD12039 (Weidenschilling 2010), suggest that "unseen" planetesimals embedded in those disks should refill the dust population as a result of their collisional evolution. The dust population of the debris disks would otherwise be depleted on a short timescale due to dust-dust collisions and other non-gravitational effects like, e.g., the Poynting-Robertson drag (see e.g. Kenyon & Bromley 2008, 2010, Weidenschilling 2010, Hahn 2011).

While we are improving our knowledge on the timescale of planetesimal accretion (see e.g. Weidenschilling 2008 for a discussion of planetary accretion in the inner and outer Solar System and Scott 2007 for constrains from meteoritic studies), the actual mechanism responsible of their formation, its efficiency and the resulting initial size-frequency distribution are still poorly known. The proposed formation scenarios differ in the assumptions on the formation environment, i.e. the Solar Nebula, and produce different size-frequency distributions of the primordial planetesimals. The standard scenario hypothesizes that the first planetesimals formed by gravitational instability of the dust in a quiescent circumsolar disk (Safronov 1969; Goldreich & Ward 1973, Weidenschilling 1980). According to Coradini, Federico & Magni (1981) and taking into account the change in density which takes place across the Snow Line, the average diameters of such planetesimals roughly span 5 - 60 km in the spatial range 1 – 40 AU or, equivalently, 5 - 25 km in the interval 1 - 10 AU. Alternative scenarios have recently been proposed, where planetesimal formation takes place in turbulent circumsolar disks (see Johansen et al. 2007, Cuzzi et al. 2008 for details on the two proposed mechanisms). Models of planetesimal formation in turbulent disks predict primordial bodies whose average sizes are orders of magnitude greater than those formed in quiescent disks. According to Chambers (2010), the "turbulent concentration" mechanism proposed by Cuzzi et al. (2008) would produce planetesimals with diameters spanning roughly 50-400 km in the radial interval 1 - 10 AU. A similar exploration performed by Cuzzi et al. (2010) shows that the outcome of such formation scenario is extremely dependent on the physical parameters of the Solar Nebula, especially on the gas density, the kinematic viscosity, the dust-to-gas ratio and the local pressure gradient. According to Cuzzi et al. (2010), the initial size-frequency distribution for the optimal set of parameters for the Solar Nebula could span the range 20-300 km at 2.5 AU and 10-100 km at 30 AU.

An attempt to constrain the formation mechanism of primordial planetesimals is the one made by Morbidelli et al. (2009). These authors did not explore a specific model of planetesimal formation in quiescent or turbulent disks but tried to estimate the primordial size-frequency distribution of the planetesimals using the present structure of the Main Asteroid Belt and the knowledge of the evolution of the Primordial and Modern Solar System and of the physical processes involved in planetary accretion. Their results suggest that the best match with the present-day size-frequency distribution of the Main Asteroid Belt is obtained for planetesimal sizes initially spanning 100 - 1000 km, a range consistent with their formation in a turbulent nebula. However, preliminary results of numerical simulations of the accretion and the collisional evolution of planetesimals in a quiescent disk considering the radial migration due to gas drag (Weidenschilling 2010) point in the opposite direction. According to Weidenschilling (2010), in fact, it is possible to reproduce the present features of the Main Asteroid Belt starting from a population of planetesimals with initial sizes of the order of 100 m. The results by Morbidelli et al. (2009) and by Weidenschilling (2010) thus suggest that, while turbulent and quiescent disks can produce very different size distributions of planetesimals, the collisional evolution of these first populations tend to produce a similar size-frequency distribution on a few Ma timescale.

### 2.1.2 *Thermal evolution of Vesta and Ceres*

Thermal evolution of Vesta has been studied by several authors, with the idea of explaining the processes that generated the differentiation of primitive bodies at the beginning of the Solar system. The idea of pristine differentiation is supported by the assumed link between Vesta and HED meteorites. Following Pieters et al. (2006) we can say that the link between Vesta and the HEDs still needs to be confirmed, even if there are several facts (Lazzaro 2009) supporting this hypothesis: first, the oxygen isotope data of the bulk of the HED suite indicate an unique origin for these meteorites (Greenwood et al. 2005, but see also the chapter by McSween et al. 2010 for a more detailed discussion). According to the analysis of Greenwood et al. (2005), these meteorite suites formed during early, global-scale melting (≥ 50 per cent) events. The second reason is that a family of asteroids dynamically linked to Vesta has been identified (Williams 1989; Zappalà et al. 1990). The members of the family, called Vestoids, seem to have a surface composition similar to Vesta (Binzel & Xu 1993). The large impact basin discovered on Vesta (Thomas et al. 1997) stimulated the idea that this large crater could be the source of Vestoids. The Vestoids could have been spread, not forming a specific dynamic family due to Yarkovsky effect (Farinella & Vokrouhlicky 1999) and to the mean motion and secular resonances able to transport fragments to near-Earth orbits (Marzari et al. 1996; Migliorini et al. 1997). Recent ground based measurements on Vestoids by De Sanctis et al. (2010) seem to indicate that there are mineralogical differences between Vestoids and Vesta. The observed differences are attributed by the authors to the variegation of the Vestian surface or to the displacement of material excavated from different layers, and not to their possible origins from different parent bodies (ibid). Also these measurements support the idea that Vesta is deeply differentiated.

Globally, the available data support the idea that differentiation should have been an important process in the initial phases of planetary formation, as discussed in Sect. 2.1.1 and, more exhaustively, by Scott 2007. Primordial differentiation of planetesimals needs a source of energy that can act efficiently on bodies with the size of Vesta or smaller: the only one that currently seems to be able to meet all the temporal and energetic requirements is the heat produced by the decay of short lived radioactive isotopes. Among the possible isotopes (see e.g. Zinner 2003 and Scott 2007) the more effective both in terms of concentration and of the production of energy seems to be $^{26}Al$. In this kind of calculations the critical parameter is the delay time $\Delta t_d$ between the injection of $^{26}$Al in the Solar Nebula and the formation of Vesta. This delay time determines the content of radioactive material at the onset of the thermal evolution of Vesta (see e.g. [Federico el al. 2011](#)). Several attempts were made in order to model the thermal evolution of Vesta: we refer the readers to [Keil (2002)](#) and to the chapter by [Zuber et al. (2011)](#) for a detailed discussion. Based on geochemical and mineralogical observations, generally it is assumed that Vesta underwent an extensive differentiation, and that was completely melted and homogenized. This brings to the paradigm that Vesta was - at the beginning of its history - characterized by the presence of an extensive magma ocean. However, recent data on HEDs seem to indicate that Vesta wasn't completely homogenized during the differentiation (see the chapter by [Mc Sween et al. 2010](#) for a discussion). [Federico et al. (2011)](#) explore different scenarios for the formation and thermal evolution of Vesta and in particular consider different delay times for the incorporation of short lived radioactive elements into the asteroid. The calculations of [Federico et al. (2011)](#) differ from those of [Ghosh & Mc Sween (1998)](#) since the ability of molten iron alloys, over a percolation threshold of only 5% in volume, is taken into account ([Walter & Tronnes, 2004](#)). [Federico et al. (2011)](#) consider also the chemical differentiation of the body due to the affinity of $^{26}$Al with silicates, which avoid to assumption of an extensive silicate melting for the differentiation of the core. In all the scenarios explored by [Federico et al. (2011)](#) the core of Vesta is formed in a temporal interval of about 1.50 Ma. This is in agreement with [Chaussidon & Gounelle (2007)](#), who confirmed a very early core formation on the basis of more recent geochemical investigations. According to [Federico et al. (2011)](#), at the beginning Vesta is heated by decaying $^{26}$Al while maintaining an almost uniform temperature from the center to a depth of about 20 km from the surface, where a steep gradient develops to reach the surface temperature of 200°K, imposed by these authors as a boundary condition due to the high efficiency of energy irradiation. Therefore, also in the first million years of its life Vesta could have had a solid crust. After a temporal interval of about 2.5 Ma, the melting temperature of both metallic and silicatic component is exceeded (ibid). By looking at the thermal profile, the almost linear decrease in temperature moving toward the surface determines a layer which can be identified with the crustal thickness of Vesta. In the model that the authors consider more realistic, the maximum temperature does not exceed 1600°K (ibid). After about 7 Ma from the formation of Vesta, a cooling phase begins and crystallization can occur. Following [Righter and Drake (1997)](Righter and Drake (1997)), the required temperatures necessary to crystallize phases present in diogenites are lower than 1783°K. As a consequence, very high temperatures inside Vesta seem to be unnecessary. The cooling rate, being about 10 K/Ma, is near the lower limit values obtained by [Akridge et al. (1998)](Akridge et al. (1998)) for H-chondrites. Eventually, a long cooling phase starts at T=16 Ma without being affected by the heat generated by the decay of long-lived radionuclides ([Federico et al. 2011](Federico et al. 2011)). According to these calculations, the attainment of a quasi-liquid structure of Vesta seems to be unrealistic.

Following the previous discussion and the timescale of the processes acting in the Solar Nebula we discussed in [Sect. 2.1.1](Sect. 2.1.1), we can state two facts that are of paramount importance for the investigation of the record of the processes which acted at the beginning Solar System:

1. the initial thermal evolution of Vesta overlaps with the temporal interval across which Jupiter is forming (see [Sect. 2.1](Sect. 2.1));

2. The crust of Vesta probably never totally melted - even if a large mantle is evolving underneath it.

As a consequence, we can hope to find on the surface of Vesta some records of the primordial bombardment that was generated on the Main Asteroid Belt by the formation of Jupiter. As such, Vesta represents a unique opportunity to investigate a temporal interval in the lifetime of the Solar System otherwise not accessible with the study of other planetary bodies.

The case of Ceres is more difficult to assess, since we do not possess meteoritic constrains on the timescale of its formation and internal evolution. Thermal models of Ceres are mainly constrained by the gravitationally relaxed shape of the dwarf planet ([Thomas et al. 2005](Thomas et al. 2005)). It is generally assumed that Ceres is a differentiated body ([McCord & Sotin 2005, Castillo-Rogez & McCord 2010](McCord & Sotin 2005, Castillo-Rogez & McCord 2010)): in such scenario, the crust of Ceres is expected to be gravitationally unstable over the liquid mantle and to have foundered repeatedly during the lifetime of the dwarf planet (see chapter by McCord et al. 2011). Recently, it has been suggested that the shape and spectral data on the dwarf planet could be also consistent with Ceres being undifferentiated and its composition being dominated by porous hydrated silicates (Zolotov 2009). In such scenario, Ceres would have formed later than in those scenarios where it underwent differentiation (ibid). It must be noted, however, that such scenario does not consider the effects of the thermal evolution of Ceres on its internal structure (McCord et al. 2011). We refer the interested readers to the chapter by McCord et al. (2011) for an in-depth discussion on the origin and evolution of Ceres.

### *2.1.3 The Jovian Early Bombardment*

Alongside the planetesimals, as we previously mentioned, the giant planets were forming across the Solar Nebula phase. It has been suggested (Turrini, Magni & Coradini 2011) that the formation of Jupiter, likely the first giant planet to form due to its greater mass and its inner position respect to the other giant planets, triggered a phase of primordial bombardment due to its rapid mass increase and its possible inward radial migration. The duration of this Jovian Early Bombardment across the Main Asteroid Belt is estimated of the order of 1 Ma (ibid). However, if Saturn formed its core and started to accrete its gaseous envelope across this timespan, its perturbations would add to those of Jupiter both in the inner and the outer Solar System. This would create a new generation of impactors and enhance both the duration and the intensity of the bombardment, therefore causing a Primordial Heavy Bombardment. The Jovian Early Bombardment and its Saturn-enhanced counterpart, the Primordial Heavy Bombardment, are likely the most violent bombardment event in the history of the Solar System, since the population of planetesimals populating the Solar Nebula had not undergone any depletion process by this time.

Across the Jovian Early Bombardment and, by extension, the Primordial Heavy Bombardment, Vesta and Ceres would have undergone impacts from both rocky bodies formed in the inner Solar System and volatile-rich bodies from the outer Solar System (ibid). According to Turrini, Magni & Coradini (2011), the intensity of the Jovian Early Bombardment depends both on the extent of Jupiter's radial migration due to disk-planet interactions and on the size-frequency distribution of the planetesimals populating the Solar Nebula. Following the results of these authors, a Solar Nebula whose population of planetesimals was dominated by large bodies (i.e. whose diameters is of the order of 100 km or greater) or where Jupiter's radial migration exceeded a few tenths of AU would have proved an extremely hostile environment for the survival of Vesta, Ceres and similarly sized bodies (ibid).

| *Jovian Early Bombardment on Vesta* | | | | | | |
|---|---|---|---|---|---|---|
| **Migration Scenario** | **Excavation Depth** | | | | | |
| | ISS impactors (Coradini et al. 1981) | OSS impactors (Coradini et al. 1981) | Primordial ISS Impactors (Morbidelli et al. 2009) | Evolved ISS Impactors (Morbidelli et al. 2009) | ISS impactors (Chambers 2010) | OSS impactors (Chambers 2010) |
| **No Migration** | 5.32 km | 10.66 km | Ablation | 9.85 km | 55.88 km | Ablation |
| **0.25 AU** | 8.55 km | 1.19 km | Ablation | 14.12 km | 46.13 km | 16.69 km |
| **0.50 AU** | 36.53 km | 0.49 km | Ablation | Ablation | Ablation | 11.62 km |
| **1.00 AU** | Ablation | 0.85 km | Ablation | Ablation | Ablation | 15.94 km |

Table 1: Collisional erosion of Vesta due to primordial planetesimals formed in quiescent (Coradini, Federico & Magni 1981) and turbulent disks (Morbidelli et al. 2009; Chambers 2010) and due to collisionally evolved planetesimals (Morbidelli et al. 2009) in the four migration scenarios of Jupiter considered by Turrini, Magni & Coradini (2011). The *ISS* and *OSS* labels indicate respectively the rocky impactors from the inner Solar System and volatile-rich impactors from the outer Solar System. The excavated depth is estimated assuming that the final radius of Vesta should be the present one. The label *Ablation* indicates those cases where the excavated volume is greater than the present volume of Vesta. Table adapted from Turrini, Magni & Coradini (2011).

The two targets of Dawn mission, Vesta and Ceres, hold a special place concerning the investigation of all these processes which took place in the Solar Nebula, if they formed at about their present locations. Vesta represents a unique probe into the timescales of the processes of planetary accretion and differentiation due to its possible link with the HED meteorites (see also McSween et al., 2010 for a more detail discussion). This connection, if confirmed, would imply that the asteroid is differentiated (see e.g. Drake, 2001; Keil, 2002, and references therein and the chapters by McSween et al. 2010 and Zuber et al. 2011). Moreover, the $^{40}Ar$-$^{39}Ar$ ages of the oldest HED meteorites (see Keil, 2002; Scott, 2007, and references therein) suggest that this asteroid is primordial, i.e. it formed and differentiated in less than 4 Ma since the formation of CAIs. If true, Vesta would be the only known surviving primordially differentiated planetesimal whose formation would date back prior to or contemporary to the formation of the giant planets. As a consequence, Vesta might be the only Solar System object which could have kept a record of the Jovian Early Bombardment. Preliminary modeling of the thermal evolution of Vesta indicates that its mantle would have been in a molten state for several Ma (see Sect. 2.1.2), i.e. across the timespan of the Jovian Early Bombardment. According to the results of Turrini, Magni & Coradini (2011) and as shown in Table 1, the Jovian early bombardment should have excavated partially or completely the primordial crust of Vesta, thus creating fractures or generating uncompensated negative gravity anomalies. These would have caused effusive phenomena from the underlying mantle, in analogy with lunar maria, or the solidification of the exposed layer of the mantle and the formation of a new basaltic crust (ibid). The crystallization epoch of these regions on the surface of Vesta would then be directly connected to the time of formation of Jupiter (ibid).

| Migration Scenario | Jovian Early Bombardment on Ceres Excavation Depth | | | | | |
|---|---|---|---|---|---|---|
| | ISL impactors (Coradini et al. 1981) | BSL impactors (Coradini et al. 1981) | Primordial SFD (Morbidelli et al. 2009) | Evolved SFD (Morbidelli et al. 2009) | ISL impactors (Chambers 2010) | BSL impactors (Chambers 2010) |
| No Migration | 13.80 km | 9.70 km | Ablation | 69.54 km | Ablation | Ablation |
| 0.25 AU | 31.45 km | 2.81 km | Ablation | Ablation | Ablation | 68.54 km |
| 0.50 AU | 61.91 km | 0.74 km | Ablation | Ablation | Ablation | 21.28 km |
| 1.00 AU | 109.60 km | 7.19 km | Ablation | Ablation | Ablation | Ablation |

Table 2: Collisional erosion of Ceres due to primordial planetesimals formed in quiescent (Coradini, Federico & Magni 1981) and turbulent disks (Morbidelli et al. 2009; Chambers 2010) and due to collisionally evolved planetesimals (Morbidelli et al. 2009) in the four migration scenarios of Jupiter considered by Turrini, Magni & Coradini (2011). The *ISS* and *OSS* labels indicate respectively the rocky impactors from the inner Solar System and volatile-rich impactors from the outer Solar System. The excavated depth is estimated assuming that the final radius of Ceres should be the present one. The label *Ablation* indicates those cases where the excavated volume is greater than the present volume of Ceres. Table adapted from Turrini, Magni & Coradini (2011).

The same arguments do not apply in the case of Ceres, since we have little constraints on the timescale of accretion and differentiation of the dwarf planet. Moreover, thermal models of the evolution of Ceres suggest that its crust could have reformed several times across the lifetime of the Solar System (see chapter by [McCord et al. 2011](#) and [Sect. 2.1.2](#)). However, Ceres is of particular interest for studying the evolution of the Solar Nebula for a different reason. According to [Turrini, Magni & Coradini (2011)](#) and as shown in Table 2, the Jovian Early Bombardment would have been more intense on Ceres than on Vesta. This is due to the combination of two factors: the bigger cross-section of Ceres with respect to Vesta and its orbit being located between the 3:1 and the 2:1 mean motion resonances with Jupiter, the main sources of impactors during the Jovian Early Bombardment (ibid). The survival of Ceres (see Table 2) is therefore even more sensitive to both Jupiter's migration and the size-distribution of the planetesimals in the disk (ibid). As a consequence, if Ceres formed near its present orbital position prior to Jupiter's formation, its survival to the Jovian Early Bombardment could be used to constrain them both and rule out implausible scenarios for the evolution of the Solar Nebula.

## *2.2 Primordial Solar System*

The Primordial Solar System is the second phase in the lifetime of the Solar System (see Figure 1): during this phase, the Solar System was shaping itself to its present structure through a series of dynamical and collisional processes which culminated with the Late Heavy Bombardment. The beginning of the Primordial Solar System can be identified with the dispersal of the nebular gas from the Solar Nebula, which is likely to have taken place earlier than 10 Ma after the condensation of CAIs. The Primordial Solar System phase ended somewhere between 3.8-4.0 Ga ago and the time of the transition to the Modern Solar System phase can be identified with that of the Late Heavy Bombardment (LHB in the following). Across this $\Delta T < 1$ Ga timespan, the evolving structure of the Solar System was extremely different from its present one. In the inner Solar System the terrestrial planets were forming from a disk of planetesimals and planetary embryos. The combined gravitational perturbations of the planetary embryos embedded in the disk and of the giant planets Jupiter and Saturn possibly triggered a rapid process of mass depletion in the orbital region of the Main Asteroid Belt (Wetherill 1992; Chambers & Wetherill 2001; Petit, Morbidelli & Chambers 2001; O'Brien, Morbidelli & Bottke 2007). In the outer Solar System it has been hypothesized that the giant planets were on a more compact orbital configuration than their present one, its radial extension spanning about 5-15 AU, and that a massive outer disk of planetesimals existed beyond the orbit of the outermost giant planet (Gomes et al. 2005; Tsiganis et al. 2005; Morbidelli et al. 2005). Planetesimal-driven migration of the giant planets would have triggered a violent phase of chaotic rearrangement of their orbits which ended with the Solar System in a configuration similar to its present one (ibid). The migration of the giant planets caused orbital resonances to sweep through the inner Solar System, destabilizing a significant fraction of the asteroids and causing the Late Heavy Bombardment (ibid).

The comparison between the reconstructed mass of the so-called "Minimum Mass Solar Nebula" (Weidenschilling 1977) and the estimated mass of the Main Asteroid Belt suggests that the orbital region between the orbit of the Earth and that of Jupiter is presently depleted in mass by about a factor 300 with respect to the beginning of the Solar Nebula phase (ibid). Such results are in agreement with those obtained by studies of the collisional evolution of the Main Asteroid Belt. By using collisional simulations, Bottke et al. (2005a) showed in fact that the present size-frequency distribution of the Main Belt is stable against collisional evolution over a timescale of Ga. According to these authors, the asteroids populating the Main Asteroid Belt should have been about 150-250 times more numerous than the present population in order to produce the present day population in the lifetime of the Solar System (ibid).

To explain the fate of the missing mass, Wetherill (1992) suggested that, at the beginning of the Primordial Solar System phase, Moon-sized to Mars-sized planetary embryos populated the inner Solar System. These planetary embryos also accounted for a significant fraction of the original mass existing in the orbital region presently occupied by the Main Asteroid Belt. The combined gravitational perturbations of the planetary embryos and of the giant planets Jupiter and Saturn would have influenced the evolution of the planetesimals in the Solar System and caused depletion in mass (ibid) with respect to that hypothesized to originally reside in the Main Asteroid Belt (Weidenschilling 1977). From a physical point of view, the early phase of this depletion process would overlap to the Jovian Early Bombardment (Turrini, Magni & Coradini 2011), yet to date all the studies of this dynamical depletion considered either Jupiter and Saturn as fully formed (Wetherill 1992; Chambers & Wetherill 2001; O'Brien, Morbidelli & Bottke 2007) or introduced them through a step-like transition (Petit, Morbidelli & Chambers 2001). The results by Wetherill (1992), Chambers & Wetherill (2001), Petit, Morbidelli & Chambers (2001) and O'Brien, Morbidelli & Bottke (2007) indicate that this "native planetary embryos" scenario would reduce the population of planetesimals in the early asteroid belt by about a factor 100 on a timescale of $10^8$ years. The planetary embryos themselves would be removed on a $\sim 10^7$ years-long timescale by being ejected from the Solar System or being accreted by planetary bodies (Petit, Morbidelli & Chambers 2001; O'Brien, Morbidelli & Bottke 2007). According to the results of Petit, Morbidelli & Chambers (2001), a few percent of the removed mass would be incorporated in bigger bodies not dynamically removed from the inner Solar System. In the simulations performed by O'Brien, Morbidelli & Bottke (2007) a significant fraction of the lost planetesimals (10-20%) were incorporated into planetary bodies due the effects of dynamical friction between planetesimals and planetary embryos. Across the timespan considered, the population of planetesimals was 1-2 orders of magnitude higher than the present one (ibid), therefore this result also implies a phase of enhanced collisional evolution of the Main Asteroid Belt, coherently with the results described by Bottke et al (2005a).

Bottke et al. (2005b) extended their previous investigation by including dynamical depletion factors due to the Yarkovsky effect and to the combined gravitational perturbations of Jupiter, Saturn and a population of primordial planetary embryos. The depletion due to the gravitational perturbations was estimated based on the depletion rate described by Petit, Morbidelli and Chambers (2001), which is higher than the ones reported by O'Brien, Morbidelli & Bottke (2007). These refined results suggested that the Main Belt suffered about a factor 150 depletion and that the best fits to the actual asteroid population are obtained for Jupiter forming in less than or about 4 Ma. These results and the estimated collisional lifetimes of the asteroids would also suggest that the population of bodies bigger than 110 km is primordial, i.e. that these bodies did not form from the break-up of bigger parent bodies. In their simulations, however, Bottke et al. (2005b) did not take into account the possible effects of the orbital rearrangement of the giant planets predicted by the Nice Model. In this scenario, the population of the Main Asteroid Belt would be further depleted by about a factor 10 across the LHB (Gomes et al. 2005). As Bottke et al. (2005b) pointed out, the implications of this possible second depletion event for the one described by the "native planetary embryos" scenario are still to be assessed.

In addition to the mass, the present orbital structure of the Main Asteroid Belt was also shaped across the Primordial Solar System phase. As discussed by O'Brien, Morbidelli & Bottke (2007) based on the results described by Levison et al. (2001), the sweeping of the resonances due to the migration of the giant planets would have completely destabilized the Main Asteroid Belt if the orbits of the asteroids were not already dynamically excited. The necessary primordial excitation, however, would be a natural consequence of the "native planetary embryos" scenario (O'Brien, Morbidelli & Bottke 2007). The migration of the giant planets, which has been suggested to be responsible for the LHB, then shaped the Main Asteroid Belt to its present orbital structure. Minton and Malhotra (2009) explored the possibility to reproduce the present orbital structure of the Main Asteroid Belt through the cumulative effects of secular perturbations and the sweeping of the resonances during a planetesimal-driven migration of the giant planets. According to these authors, such scenario could reproduce well the observational constrains (ibid). However, Morbidelli et al. (2010) argued that the migration rate assumed by Minton and Malhotra (2009) are too fast for such kind of scenario and that more realistic migration rates fail to reproduce orbital structures compatible with that of the Main Asteroid Belt.

Similar results are obtained by Walsh & Morbidelli (2011) in exploring the effects of a possible early planetesimal-driven migration of the giant planets. According to Morbidelli et al. (2010), the observational constrains can be better reproduced either if the migration followed a path similar to the one described in the Nice Model (Gomes et al. 2005, Tsiganis et al. 2005) or an even more drastic "Jumping Jupiters" migration pattern like those proposed to explain the peculiar orbital structures of several multi-planet extrasolar systems (Weidenschilling & Marzari 1996). Following the results of Morbidelli et al. (2010), a "Jumping Jupiters" scenario would reduce the extent of the depletion event associated to the LHB from the factor 10 estimated by Gomes et al. (2005) to about a factor 2 and would leave the inclination distribution of the asteroids essentially unchanged.

Vesta and Ceres both play a fundamental role in constraining the evolution of the Primordial Solar System and the transition to the Modern Solar System, as also discussed in the chapter by O'Brien & Sykes (2011). Ceres, being the most massive object which survived to present time in the Main Asteroid Belt, represents an important probe of the efficiency of the depletion mechanism which operated in that orbital region across the Primordial Solar System phase. Vesta, with its primordial basaltic crust, is the only body we know of that can constrain the collisional evolution of the Main Asteroid Belt across the Primordial Solar System phase and shed new light on the events that characterized this ancient and violent time. The knowledge we can gather on the evolution of the Primordial Solar System with Dawn will be fundamental to better understand those features of Vesta and Ceres that are a legacy of their formation histories.

## *2.3 Modern Solar System*

While the most violent part of the history of the Solar System ended with the transition to the Modern Solar System phase (see Figure 1), the collisional evolution of Vesta and Ceres did not end with the LHB. We will limit here to mention the processes acting across this 3 Ga long temporal interval, referring the readers to the chapter by O'Brien and Sykes (2011) for a more detailed discussion.

The violent depletion processes acting in the Primordial Solar System left place to slow, secular depletion mechanisms. Minton & Malhotra (2010) estimated that, since the Main Asteroid Belt reached its present configuration, chaotic diffusion into the resonances with the giant planets could have caused a depletion of a factor 2 of the population of large asteroids (D > 10-30 km). The population of small asteroids likely suffered a higher secular depletion, due to the effects of Yarkovsky-driven diffusion (see e.g. O'Brien & Sykes 2011; Asphaug 2009, and references therein). A fraction of these diffused asteroids likely impacted on Vesta and Ceres, thus contributing to their collisional evolution (see O'Brien & Sykes 2011, and references therein). While individually less frequent, impacts across the Modern Solar System phase cumulatively accounted for a significant part of the collisional histories of Vesta and Ceres (see e.g. Bottke et al. 2005b).

Even if less frequent, such impacts would be characterised by average velocities higher than those of their counterparts which took place in the previous phases of the life of the Solar System. In particular, it has been suggested that the impact of an asteroid with diameter of about 30 km could have been responsible for the formation of the Vestoids less than 1 Ga ago, basing on dynamical constrains (Marzari et al. 1996). Similarly, the cosmic-ray exposure ages of HED meteorites suggest the possibility they were ejected from Vesta by recent impacts, i.e. less than 100 Ma ago: we refer the readers to the chapter by McSween et al. (2010) for further details.

The secular evolution of the Main Asteroid Belt across the Modern Solar System phase likely played an important role in shaping the present appearances of Vesta and Ceres. To explore the ancient past of the Solar System through Dawn mission, we need to take into account the effects of this last long temporal interval and how they could have affected the signature of the formation and evolution histories of Vesta and Ceres.

## 3. Concluding remarks

As we discussed, Vesta and Ceres were characterized by a thermal evolution driven by the decay of short-lived radionuclides that were present or injected in the Solar Nebula. The thermal evolution of Vesta took place contemporary to the formation of the giant planets in the Solar Nebula phase, while we have little constrains on the timescale of the thermal evolution of Ceres. After they differentiated, Vesta and Ceres crossed the whole history of the Solar System, surviving to the violent dynamical and collisional processes that characterized its early stages. These occurrences pose Vesta and Ceres in a unique position among the bodies populating the inner Solar System.

Through the data that Dawn mission will collect, Vesta and Ceres will play a fundamental role in unveiling the origin and the evolution of the Solar System as we know it. As discussed also by O'Brien & Sykes (2011), the different phases of the history of the Solar System would have left different signatures on the surfaces and the structures of these two bodies. While disentangling the marks of 4.5 Ga of thermal and collisional evolution will not be an easy task, through Dawn mission we will get a better insight on a number of important unsolved problems. Between those, the missing mass of the Main Asteroid Belt, the timescale of formation and differentiation of the primordial planetesimals, the timescale of formation of Jupiter and the primordial and late migration of the giant planets.

## Acknowledgements

The authors wish to thank the two anonymous referees for their comments on the manuscript. D.T. wishes to thank also Romolo Politi and Roberto Peron for their comments and suggestions on the manuscript. This research has been supported by the Italian Space Agency (ASI) through the ASI-INAF contract I/015/07/0.

## Bibliography

1. Amelin Y., Krott A. N., Hutcheon I. D., Ulyanov A. A., Lead Isotopic Ages of Chondrules and Calcium-Aluminum-Rich Inclusions, 2002, Science, 297, 1678-1683.
2. Amelin Y., Kaltenbach A., Iizuka T., Stirling C. H., Ireland T. R., Petaev M., Jacobsen S. B., U-Pb chronology of the Solar System's oldest solids with variable $^{238}U/^{235}U$, Earth and Planetary Science Letters, 2010, 300, 343-350.
3. Akridge G., Benoit P. H., Sears D. W. G., Regolith and megaregolith formation of H chondrites: Thermal constraints on the parent body, 1998, Icarus, 132, 185-195.
4. Asphaug E., Growth and Evolution of Asteroids, 2009, Annual Reviews of Earth and Planetary Sciences, 37, 413-448

5. Baker J. A., Bizzarro M., Wittig N., Connelly J. N., Haack H., Early planetesimal melting from an age of 4.5662 Gyr for differentiated meteorites, 2005, Nature, 436, 1127-1131.

6. Bertotti B., Farinella P., Vokrouhlický D., Physics of the Solar System: dynamics and evolution, space physics, and spacetime structure, 2003, The Netherlands, Kluwer Academic Publishers, ISBN: 1-4020-1428-7.

7. Binzel R. P., Xu, S., Chips off of asteroid 4 Vesta: Evidence for the parent body of basaltic achondrite meteorites, 1993, Science, 260, 186-191.

8. Bizzarro M., Baker J. A., Haack H., Luundgard K. L., Rapid timescales for accretion and melting of differentiated planetesimals inferred from 26Al-26Mg chronometry, 2005, The Astrophysical Journal, 632, L41-L44.

9. Bottke W. F., Durda D. D., Nesvorny D., Jedicke R., Mordibelli A., Vokrouhlicky D., Levison H., The fossilized size distribution of the main asteroid belt, 2005, Icarus, 175, 111-140.

10. Bottke W. F., Durda D. D., Nesvorny D., Jedicke R., Mordibelli A., Vokrouhlicky D., Levison H., Linking the collisional history of the main asteroid belt to its dynamical excitation and depletion, 2005, Icarus, 179, 63-94.

11. Bouvier A., Wadhwa M., The age of the Solar System redefined by the oldest Pb–Pb age of a meteoritic inclusion, 2010, Nature Geoscience, 3, 637 – 641.

12. Castillo-Rogez J., McCord T. B., Ceres' evolution and present state constrained by shape data, 2010, Icarus, 205, 443-459.

13. Chambers J. E., Planetesimal formation by turbulent concentration, 2010, Icarus, 208, 505-517.

14. Chambers J. E., Wetherill G. W., Planets in the asteroid belt, 2001, Meteoritics and Planetary Science, 36, 381-399.

15. Chaudisson M., Gounelle M., Short-lived radioactive nuclides in meteorites and early solar system process, 2007, Comptes Rendus Geosciences, 339, 872-884.

16. Connelly J. N., Amelin Y., Krot A. N., Bizzarro M., Chronology of the Solar System's oldest solids, 2008, The Astrophysical Journal, 675, L121-L124.

17. Coradini A., Federico C., Magni G., Formation of Planetesimals in an evolving Protoplanetary Disk, 1981, Astronomy and Astrophysics, 98, 173-185.

18. Coradini A., Magni G., Turrini D., From Gas to Satellitesimals: Disk Formation and Evolution, 2010, Space Science Reviews, 153, 411-429.

19. Cuzzi J. N., Hogan R. C., Shariff K., Toward planetesimals: Dense chondrule clumps in the protoplanetary nebula, 2008, The Astrophysical Journal, 687, 1432-1447.

20. Cuzzi J. N., Hogan R. C., Bottke W. F., Towards initial mass functions for asteroids and Kuiper Belt Objects, 2010, Icarus, 208, 518-538.

21. De Pater I., Lissauer J. J., Planetary Sciences, 2001, Cambridge (UK), Cambridge University Press, ISBN: 0521482194.

22. De Sanctis M. C., Ammannito E., Migliorini A., Lazzaro D., Capria M. T., McFadden L., Mineralogical characterization of some V-type asteroids, in support of the NASA Dawn mission, 2011, Monthly Notices of the Royal Astronomical Society, accepted for publication.
23. Drake M. J., The eucrite/Vesta story, 2001, Meteoritics and Planetary Science, 36, 501-513.
24. Farinella P., Vokrouhlicky Y. D., Semimajor axis mobility of asteroidal fragments, 1999. Science 283, 1507-1510.
25. Federico C., Coradini A., Pauselli C., Vesta Thermal and Structural Evolution Models, 2011, submitted to Planetary and Space Science.
26. Ghosh A., McSween H. Y., Extended Thermal History (100 MA Long) for Asteroid 4 Vesta Based on Radionuclide and Collisional Heating, 1996, Lunar and Planetary Science, volume 27, page 407.
27. Ghosh A., McSween H. Y., A Thermal Model for the Differentiation of Asteroid 4 Vesta, Based on Radiogenic Heating, 1998, Icarus, 134, 187-206.
28. Goldreich P., Ward W. R., The formation of planetesimals, 1973, The Astrophysical Journal, 183, 1051-1062.
29. Gomes R., Levison H.F., Tsiganis K., Morbidelli A., Origin of the cataclysmic Late Heavy Bombardment period of the terrestrial planets, Nature, 2005, 435, 466-469.
30. Greenwood R. C., Franchi I. A., Jambon A., Buchanan P. C., Widespread Magma Oceans On Asteroidal Bodies In The Early Solar System, 2005, Nature, 435, 916-918.
31. Hahn J. M., Diagnosing Circumstellar Debris Disks, 2010, The Astrophysical Journal, 719, 1699-1714.
32. Haisch K. E., Lada E. A., Lada C. J., Disk frequencies and lifetimes in young clusters, 2001, The Astrophysical Journal, 553, L153-L156.
33. Jayawardhana R. J., Coffey A., Scholz A., Brandeker A., van Kerkwijk M. H., Accretion Disks around Young Stars: Lifetimes, Disk Locking, and Variability, 2006, The Astrophysical Journal, 648, 1206–1218.
34. Johansen A., Oishi J. S., Mac Low M.-M., Klahr H., Henning T., Youdin A., Rapid planetesimal formation in turbulent circumstellar disks, 2007, Nature, 448, 1022-1025.
35. Keil K., Geological History of Asteroid 4 Vesta: The Smallest Terrestrial Planet, 2002, Asteroids III, Eds. W. F. Bottke Jr., A. Cellino, P. Paolicchi, and R. P. Binzel, University of Arizona Press, Tucson, 573-584.
36. Kenyon S. J., Bromley B. C., Dusty Rings: Signposts of Recent Planet Formation, 2002, The Astrophysical Journal, 577, L35-L38.
37. Kenyon S. J., Bromley B. C., Variations on Debris Disks: Icy Planet Formation at 30-150 AU for 1-3 $M_{solar}$ Main-Sequence Stars, 2008, The Astrophysical Journal Supplement Series, 179, 451-483.

38. Kenyon S. J., Bromley B. C., Variations on Debris Disks. II. Icy Planet Formation as a Function of the Bulk Properties and Initial Sizes of Planetesimals, 2010, The Astrophysical Journal Supplement, 188, 242-279.

39. Kleine T., Münker C., Mezger K., Palme H., Rapid accretion and early core formation on asteroids and the terrestrial planets from Hf-W chronometry, 2002, Nature, 418, 952-955.

40. Kleine T., Palme H., Mezger K., Halliday A. N., Hf-W Chronometry of Lunar Metals and the Age and Early Differentiation of the Moon, 2005, Science, 310, 1671-1674.

41. Lazzaro, D., Basaltic Asteroids: A New Look On The Differentiation Process In The Main Belt, 2009, Proceedings of the XII Latin American IAU Regional Meeting, Eds. G. Magris, G. Bruzual, & L. Carigi, Revista Mexicana de Astronomía y Astrofísica (Serie de Conferencias), 35, 1-6.

42. Levison H. F., Dones L., Chapman C. R., Stern S. A., Duncan M. J., Zahnle K., Could the Lunar ``Late Heavy Bombardment'' Have Been Triggered by the Formation of Uranus and Neptune?, 2001, Icarus, 151, 286-306.

43. Lissauer J. J., Hubickyi O., D'Angelo G., Bodenheimer P., Models of Jupiter's growth incorporating thermal and hydrodynamics constraints, 2009, Icarus, 199, 338-350.

44. Lugmair G. W., Shukolyukov A., Early solar system timescales according to 53Mn-53Cr systematics, 1998, Geochimica et Cosmochimica Acta, 62, 2863-2886.

45. Markowski A., Quitté G., Halliday A. N., Kleine T., Tungsten isotopic compositions of iron meteorites: Chronological constraints vs. cosmogenic effects, 2006, Earth and Planetary Science Letters, 242, 1-15.

46. Marzari F., Cellino A., Davis D. R., Farinella P., Zappala V., Vanzani V., Origin and evolution of the Vesta asteroid family, 1996, Astronomy and Astrophysics, 316, 248-262.

47. McCord T. B., Castillo-Rogez J., Rivkin A., Ceres: Its Origin, Evolution and Structure and Dawn's Potential Contribution, 2011, Space Science Reviews, Online First, DOI: 10.1007/s11214-010-9729-9.

48. McCord T. B., Sotin C., Ceres: Evolution and current state, 2005, Journal of Geophysical Research, 110, E05009.

49. McSween H. Y., Mittlefehldt D. W., Beck A. W., Mayne R. G., McCoy T. J., HED Meteorites and Their Relationship to the Geology of Vesta and the Dawn Mission, 2010, Space Science Reviews, Online First, DOI:10.1007/s11214-010-9637-z.

50. Meyer M. R., Circumstellar Disk Evolution: Constraining Theories of Planet Formation, 2008, Proceedings of the International Astronomical Union, 4, 111-122, DOI: 10.1017/S1743921309031767.

51. Migliorini F., Morbidelli A., Zappalà V., Gladman B., Bailey M. E., Cellino A., Vesta fragments from v6 and 3:1 resonances: Implications for V-type NEAs and HED meteorites, 1997, Meteoritics and Planetary Science, 32, 903-916.

52. Minton D. A., Malhotra R., A record of planet migration in the main asteroid belt, 2009, Nature, 457, 1109-1111.
53. Minton D. A., Malhotra R., Dynamical erosion of the asteroid belt and implications for large impacts in the inner Solar System, Icarus, 207, 744-757.
54. Morbidelli A., Levison H.F., Tsiganis K., Gomes R., Chaotic capture of Jupiter's Trojan asteroids in the early Solar System, Nature, 2005, 435, 462-465.
55. Morbidelli A., Bottke W. F., Nesvorny D., Levison H. F., Asteroids were born big, 2009, Icarus, 204, 558-573.
56. Morbidelli A., Brasser R., Gomes R., Levison H. F., Tsiganis K., Evidence from the Asteroid Belt for a Violent Past Evolution of Jupiter's Orbit, 2010, The Astronomical Journal, 140, 1391-1401.
57. O'Brien D. P., Morbidelli A., Bottke W. F., The primordial excitation and clearing of the asteroid belt - Revisited, 2007, Icarus, 191, 434-452.
58. O'Brien D. P., Sykes M. V., The Asteroid Belt – Creation and Destruction of Planets, 2011, Space Science Reviews, this volume.
59. Papaloizou J. C. B., Nelson R. P., Kley W., Masset F. S., Artymowicz P., Disk-Planet Interactions During Planet Formation, 2007, Protostars and Planets V, Eds. B. Reipurth, D. Jewitt, and K. Keil, University of Arizona Press, Tucson, 655-668.
60. Petit J., Morbidelli A., Chambers J., The Primordial Excitation and Clearing of the Asteroid Belt, 2001, Icarus, 153, 338-347.
61. Pieters C., Binzel R. P., Bogard D., Hiroi T., Mittlefehldt D. W., Nyquist L., Rivkin A., Takeda H., Asteroid-meteorite links: the Vesta conundrum(s), 2006, Asteroids, Comets, Meteors Proceedings IAU Symposium No. 229, D. Lazzaro, S. Ferraz-Mello & J.A. Fernandez Eds, Cambridge, Cambridge University Press, 273-288
62. Quitté G., Birck J. L., Tungsten isotopes in eucrites revisited and the initial $^{182}Hf/^{180}Hf$ of the solar system based on iron meteorite data, 2004, Earth and Planetary Science Letters, 219, 201-207.
63. Righter K., Drake M. J., A magma ocean on Vesta: Core formation and petrogenesis of eucrites and diogenites, 1997, Meteoritics & Planetary Science, 32, 929-944.
64. Safronov V. S., Evolution of the Protoplanetary Cloud and Formation of the Earth and Planets, 1969, Nauka Press, Moscow, English Translation: NASA TTF-677.7.
65. Scott E. R. D., Meteoritics and dynamical constrains on the growth mechanisms and formation times of asteroids and Jupiter, 2006, Icarus, 185, 72-82.
66. Scott E. R. D., Chondrites and the Protoplanetary Disk, 2007, Annual Reviews of Earth and Planetary Sciences, 35, 577-620.
67. Shukolyukov A., Lugmair G. W., Live Iron-60 in the early solar system, 1993, Science, 259, 1138-1142.
68. Srinivasan G., Goswami J. N., Bhandari N., 26Al in Eucrite Piplia Kalan: Plausible Heat Source and Formation Chronology, 1999, Science, 284, 1348-1350.

69. Thomas P. C., Parker J. W., McFadden L. A., Russell C. T., Stern S. A., Sykes M. V., Young E. F., Differentiation of the asteroid Ceres as revealed by its shape, 2005, Nature, 437, 224-226.

70. Tsiganis K., Gomes R., Morbidelli A., Levison H. F., Origin of the orbital architecture of the giant planets of the Solar System, Nature, 2005, 435, 459-461.

71. Turrini D., Magni G., Coradini A., Probing the history of Solar System through the cratering records on Vesta and Ceres, 2011, Monthly Notices of the Royal Astronomical Society, Online Early, DOI: 10.1111/j.1365-2966.2011.18316.x.

72. Urey, H.C., The Cosmic Abundances of Potassium, Uranium, and Thorium and the Heat Balances of the Earth, the Moon, and Mars, 1955, Proceedings of the National Academy of Sciences of the United States of America, 41, 127-144.

73. Walsh K. J., Morbidelli A., The effect of an early planetesimal-driven migration of the giant planets on terrestrial planet formation, 2011, Astronomy and Astrophysics, 526, id.A126.

74. Walter M. J., Tronnes R. G., pp $CO_2$ Early Earth differentiation, 2004, Earth and Planetary Science Letters, 225, 253-269.

75. Weidenschilling S. J., The distribution of mass in the planetary system and solar nebula, 1977, Astrophysics and Space Science, 51, 153-158.

76. Weidenschilling S. J., Dust to planetesimals - Settling and coagulation in the solar nebula, 1980, Icarus, 44, 172-189.

77. Weidenschilling S. J., Marzari F., Gravitational scattering as a possible origin for giant planets at small stellar distances, 1996, Nature, 384, 619-621.

78. Weidenschilling S. J., Accretion of planetary embryos in the inner and outer solar system, 2008, Physica Scripta, 130, 014021.

79. Weidenschilling S. J., Collisional and luminosity evolution of a debris disk: the case of HD 12039, 2010, The Astrophysical Journal, 722, 1716-1726.

80. Weidenschilling S. J., Were Asteroids Born Big? An Alternative Scenario, 41st Lunar and Planetary Science Conference, March 1-5 2010 held in The Woodlands, Texas, LPI Contribution No. 1533, p. 1453.

81. Wetherill G. W., An alternative model for the formation of asteroids, 1992, Icarus, 100, 307-325.

82. Williams J. G., Asteroid family identifications and proper elements, 1989, in Asteroids II, Eds. R. P. Binzel, T. Gehrels, & M. S. Matthews, Tucson, University of Arizona Press, 1034-1072.

83. Yang J., Goldstein J. I., Scott E. D. R., Iron meteorite evidence for early formation and catastrophic disruption of protoplanets, 2007, Nature, 446, 888-891.

84. Yin Q., Jacobsen S. B., Yamashita K., Blichert-Toft J., Télouk P., Albarède F., A short timescale for terrestrial planet formation from Hf-W chronometry of meteorites, 2002, Nature, 418, 949-952.

85. Zappalà V., Cellino A., Farinella P., Knezevic Z., Asteroid families I - Identification by hierarchical clustering and reliability assessment, 1990, Astronomical Journal, 100, 2030-2046.
86. Zinner E., An Isotipic View of the Early Solar System, 2003, Science, 300, 265-267.
87. Zolotov M. Y., On the composition and differentiation of Ceres, 2009, Icarus, 204, 183-193.
88. Zuber et al., Vesta: Its Origin, Evolution and Structure and Dawn's Potential Contribution, 2011, in preparation.